\documentclass[runningheads]{llncs}
\usepackage[T1]{fontenc}
\usepackage{graphicx}
\usepackage{multirow}
\usepackage{mathtools}
\usepackage{amsmath,amsfonts,amssymb}
 \usepackage{multirow}

\usepackage{subcaption}
\usepackage{bm}
\usepackage{siunitx}
\usepackage{hyperref}
\hypersetup{pdftex=true, colorlinks=true, breaklinks=true, linkcolor=blue, menucolor=blue, citecolor=blue, urlcolor=blue}
\usepackage{listings}
\usepackage{xcolor}
\usepackage{comment}
\usepackage{enumitem}

\begin{document}

\title{TOaCNN: Adaptive Convolutional Neural Network for Multidisciplinary Topology Optimization}

\author{Khaish Singh Chadha \and
Prabhat Kumar}
\institute{Department of Mechanical and Aerospace Engineering, Indian Institute of Technology Hyderabad, 502285 Telangana, India\\
\email{khaish13@gmail.com},
\email{pkumar@mae.iith.ac.in}}
\maketitle 
Published\footnote{This pdf is the personal version of an article whose final publication is available at \href{https://link.springer.com/chapter/10.1007/978-981-96-1158-4_43}{Advances in Multidisciplinary Design, Analysis and Optimization}}\,\,\,in \textit{Advances in Multidisciplinary Design, Analysis and Optimization}, 
\href{https://link.springer.com/chapter/10.1007/978-981-96-1158-4_43}{DOI:10.1007/978-981-96-1158-4\_43} 
%%%%%%%%%%%%%%%%%%%%%%%%%%%%%%%%%%%%%%%%%%%%%%%%
\begin{abstract}
This paper presents an adaptive convolutional neural network (CNN) architecture that can automate diverse topology optimization (TO) problems having different underlying physics. The architecture uses the encoder-decoder networks with dense layers in the middle which includes an additional adaptive layer to capture complex geometrical features. The network is trained using the dataset obtained from the three open-source TO codes involving different physics. The robustness and success of the presented adaptive CNN are demonstrated on compliance minimization problems with constant and design-dependent loads and material bulk modulus optimization. The architecture takes the user's input of the volume fraction. It instantly generates optimized designs resembling their counterparts obtained via open-source TO codes with negligible performance and volume fraction error. 

\keywords{Topology optimization, Machine learning, Convolutional neural network,  Standard architecture}

\end{abstract}
%%%%%%%%%%%%%%%%%%%%%%%%%%%%%%%%%%%%%%%%%%%%%%%%

\section{Introduction}
Topology optimization (TO) is a computational technique that determines the efficient material distribution within a  design domain while optimizing an objective with predetermined constraints and boundary conditions~\cite{sigmund2013topology}. With the remarkable progress in additive manufacturing processes, TO's utility and demand steadily increase for various design problems~\cite{langelaar2017additive}. TO process typically involves four stages: (i) parameterizing the given design domain, (ii) conducting finite element analyses for relevant physical aspects, (iii) assessing the objective function, constraints, and their corresponding sensitivities, and (iv) iterating through the optimization procedure. TO becomes notably intricate and challenging, especially when dealing with multi-physics problems~\cite{choi2016comparison} and design-dependent loads~\cite{kumar2020topology,kumar2022topology}. These computational requirements present significant challenges and hinder the broader practical implementation of TO methods~\cite{choi2016comparison}. This paper introduces an adaptive convolutional neural network-based architecture for multidisciplinary design optimization to complement traditional TO methods. Once trained, this architecture provides optimized designs instantly for the same boundary and force conditions it was trained.

\begin{figure}[h]
	\begin{subfigure}{0.30\textwidth}
		\centering
		\includegraphics[scale=0.9]{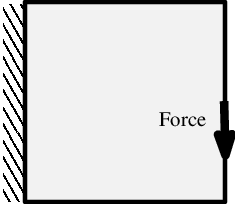}
		\caption{}
		\label{fig:cantileverDD}
	\end{subfigure}
	\quad
	\begin{subfigure}{0.30\textwidth}
		\centering
		\includegraphics[scale=0.9]{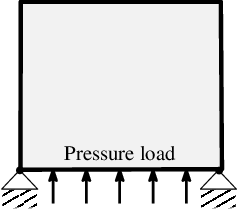}
		\caption{}
		\label{fig:archDD}
	\end{subfigure}
    \quad
 	\begin{subfigure}{0.30\textwidth}
		\centering
		\includegraphics[scale=0.9]{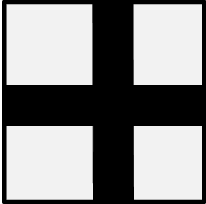}
		\caption{}
		\label{fig:microDD}
	\end{subfigure}
	\caption{Problem descriptions. (a) Cantilever beam (b) Pressure loadbearing arch structure (c) Micro-structure design } \label{fig:Deasign domains}
\end{figure}

Integration of deep learning into optimization tasks has emerged as a promising avenue~\cite{sosnovik2019neural,banga20183d,harish2020topology,chandrasekhar2021tounn,seo2023development}. TO codes generate visual representations of optimized designs in the form of images. On the other hand, Convolutional Neural Networks (CNNs) have demonstrated exceptional proficiency in extracting valuable features and discerning intricate patterns and relationships within image data. This is one of the key reasons for their adoption of automating the TO process. Previous attempts at employing CNNs for TO have shown remarkable performance~\cite{banga20183d,wang2022deep,harish2020topology}. However,  the neural network architectures developed in the above-mentioned references are tailored for a specific optimization task and lack the generalization ability to adapt to new optimization problems. To fill this gap, in this work, we propose an adaptive CNN architecture capable of automating TO problems across various problems. Efficiacy of the proposed architecture is demonstrated on three distinct TO problems with different physics: compliance minimization with constant and design-dependent loads and material bulk modulus optimization problems (Fig.~\ref{fig:Deasign domains}).

The remainder of the paper is structured as follows: Sec.~\ref{Sec:Sec2_Methodology} provides the methodology--problem description, proposed CNN neural network architecture, and generation of data. Results and discussions are presented in Sec.~\ref{Sec:Sec3_res_dis}. Lastly, Sec.~\ref{Sec:Sec4_Conclusion} outlines the concluding remarks.

\begin{figure}[h!]
\centering
    \includegraphics[scale=0.85]{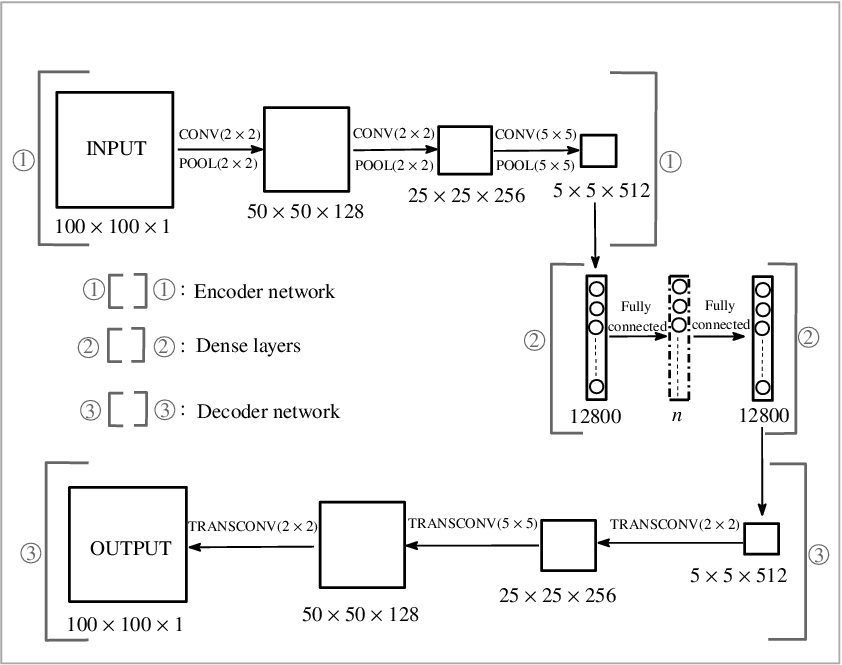}
    \caption{General Architecture of the Adaptive Convolutional Neural Network (CNN)}
    \label{fig:CNNarchitech}
\end{figure}

\section{Methodology}\label{Sec:Sec2_Methodology}
This section presents the problem description, the proposed neural network architecture, and the training data generation methodology.
\subsection{Problem description}
As mentioned above, we select a set of three problems involving different physics to train the proposed adaptive CNN architecture and obtain the optimized designs. First, the compliance minimization problem for designing a cantilever beam (Fig.~\ref{fig:cantileverDD}) is taken. The data pertaining to that is generated using \texttt{top88} code~\cite{andreassen2011efficient}. Second, the compliance minimization problem for the loadbearing arch structure with the design-dependent load (Fig.~\ref{fig:archDD}) is considered.  \texttt{TOPress} code~\cite{kumar2023TOPress} is used to generate the data for the arch problem. Design-dependent loads within a TO setting pose several challenges~\cite{kumar2020topology,kumar2022topology,kumar2023TOPress} as these loads change their direction, magnitude, and direction with TO evolution. However, once the architecture is trained, those challenges no longer exist. Third, we solve the material bulk modulus optimization problem per~\cite{xia2015design}. \texttt{topX} code~\cite{xia2015design} is used to generate the training data.
\subsection{Neural Network Architecture}
We propose a deep learning model using a convolutional neural network. The network contains an encoder-decoder architecture with dense layers added at the bottleneck region. The encoder and decoder part of the network are purely convolutional, whereas the dense layer is adaptive; one can change the number of neurons $n$ in the middle layer of the dense layers block (Fig.~\ref{fig:CNNarchitech}) as per the complexity of geometrical feature of the optimized designs. We name the middle layer ``adaptive layer.'' The architecture combines the strengths of convolutional layers for feature extraction from images and fully connected layers (dense layers) for relatively more abstract, high-level processing. The three main parts of the proposed architecture are discussed below.

\textbf{Encoder network:} It plays a prominent role in extracting meaningful information from the training data while reducing the dimensionality of the input data. Herein, the input image has a size of $(100\times100\times1)$, which is down-sampled by the encoder network to a size of $(5\times5\times512)$. This is achieved by three successive convolutional and max pooling operations (Fig.~\ref{fig:CNNarchitech}).  All the convolution operations performed are ``same'' convolutions to ensure that the information at the edges of the input image is fully considered in the output feature map. Max-pooling operations are responsible for the down-sampling of the image. The stride value of each convolution operation is kept as $(1\times1),$ whereas that for the max-pooling is kept equal to the filter size.

\textbf{Dense layers:} Output of the encoder network with $(5\times5\times512)$ is flattened out to create the first dense layer with 12800 neurons (Fig.~\ref{fig:CNNarchitech}); the optional adaptive dense layer follows this (indicated by dotted boundary in Fig.~\ref{fig:CNNarchitech}) and another dense layer having 12800 neurons (Fig.~\ref{fig:CNNarchitech}). The adaptive layer equips the network with the capability to automate a broad spectrum of optimization tasks. The choice to include the adaptive layer and determine the number of neurons in that layer should be based on the complexity of the optimized designs for the particular TO problem. The proposed CNN architecture, when devoid of the adaptive layer, is termed the \textit{Base} architecture.
 
\textbf{Decoder network:} The output of the dense layers is to form a multi-channel image having a size of  $(5\times5\times512)$, which is then up-scaled to $(100\times100\times1)$ via successive transpose convolution operations (Fig.~\ref{fig:CNNarchitech}). Filter size and number of filters used in each transpose convolution are mentioned in Fig.~\ref{fig:CNNarchitech}. Each transpose convolution layer has a stride value that equals the filter size. We provide zero padding to both input and output for all the transpose convolution operations. 

We use ``ReLU'' activation function for the convolution and transpose convolution. Mean squared error is the cost function. ``Adam'' optimizer is employed for efficient training of the neural network.

\begin{figure}[h]
	\begin{subfigure}{0.22\textwidth}
		\centering
		\includegraphics[scale=0.10]{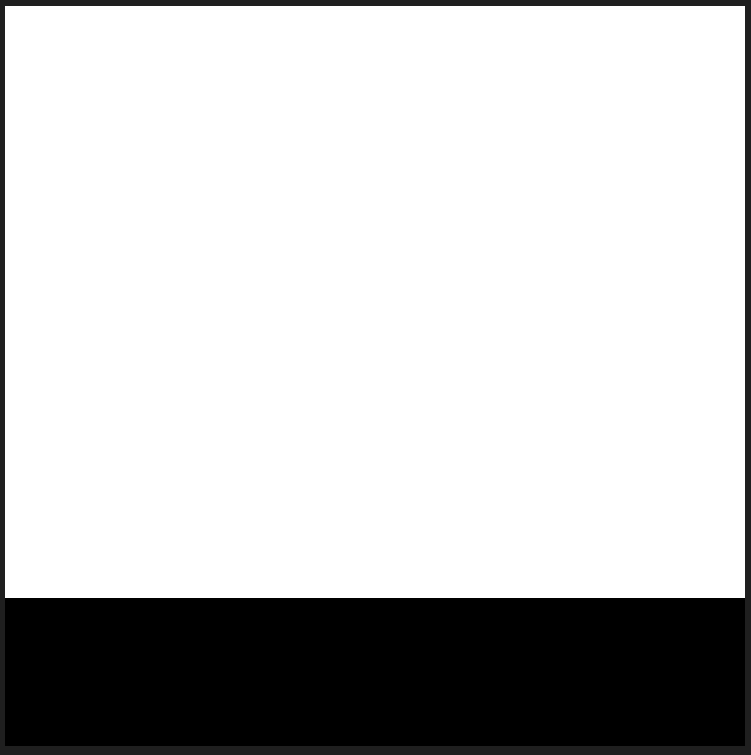}
		\caption{}
		\label{fig:inputimage}
	\end{subfigure}
\quad
	\begin{subfigure}{0.22\textwidth}
		\centering
		\includegraphics[scale=0.10]{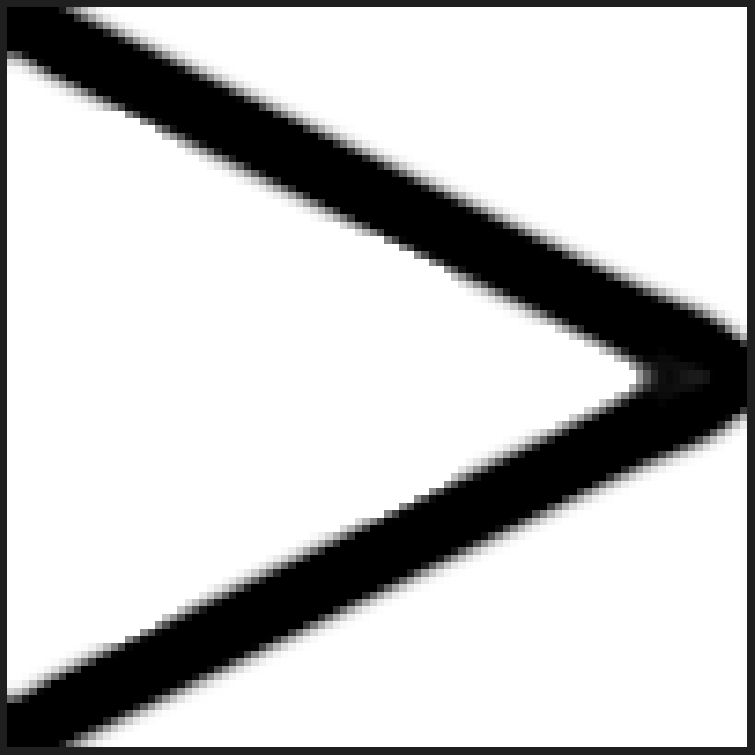}
		\caption{}
		\label{fig:88mid_in_out}
	\end{subfigure}
\quad
	\begin{subfigure}{0.22\textwidth}
		\centering
		\includegraphics[scale=0.10]{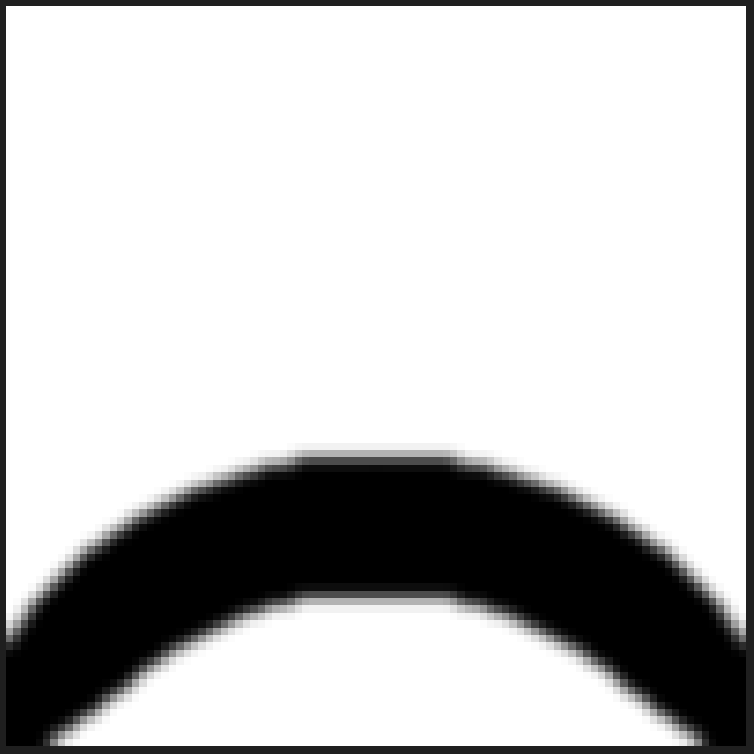}
		\caption{}
		\label{fig:press_in_out}
	\end{subfigure}
\quad
	\begin{subfigure}{0.22\textwidth}
		\centering
		\includegraphics[scale=0.10]{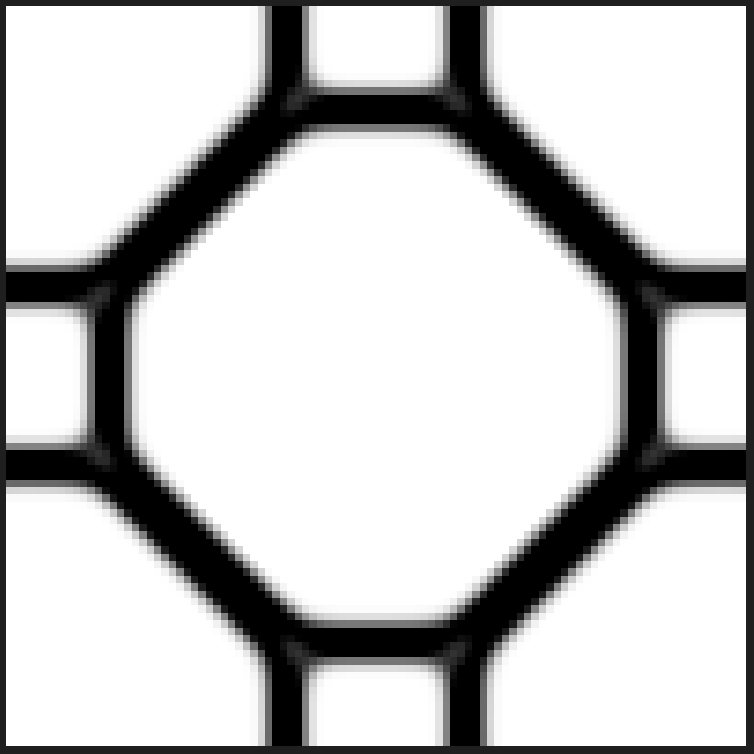}
		\caption{}
		\label{fig:x_in_out}
	\end{subfigure}
	\caption{Training data set: Input and target images. (\subref{fig:inputimage}) Input image. Target images: (\subref{fig:88mid_in_out})  Cantilever beam (Fig.~\ref{fig:cantileverDD}), (\subref{fig:press_in_out}) Loadbearing arch (Fig.~\ref{fig:archDD}) and (\subref{fig:x_in_out}) Material bulk modulus (Fig.~\ref{fig:microDD})} \label{fig:Training set}
\end{figure}

\subsection{Generation of training data}\label{Sec:GentrainingData}
The proposed architecture takes input data as images. $\texttt{top88}$~\cite{andreassen2011efficient}, $\texttt{TOPress}$~\cite{kumar2023TOPress} and $\texttt{topX}$~\cite{xia2015design} MATLAB codes are used to generate target (optimized designs, cf.~Fig.~\ref{fig:Training set}) images having size $100\times 100$ (i.e., $100\times100$ finite elements are used to parameterize the design domains) for the problems by varying the volume fraction from 0.01 to 0.95 with an increment of 0.01. We use $\{\mathtt{penal =3,\, rmin =2.4,\,ft=1}\}$, \linebreak $\{\mathtt{penal =3,\, rmin=2.4,\,etaf = 0.2,\,betaf =8,\,lst = 1,\,maxit =100}\}$ and \linebreak$\{\mathtt{penal =3,\, rmin =2.4,\,ft=1}\}$ as inputs in $\texttt{top88}$~\cite{andreassen2011efficient}, $\texttt{TOPress}$~\cite{kumar2023TOPress} and $\texttt{topX}$~\cite{xia2015design} MATLAB codes for the respective problems (Fig.~\ref{fig:Deasign domains}). For the input images, we produce black-and-white images wherein the black pixels in the images are equal to the volume fractions (Fig.~\ref{fig:inputimage}). If generating data within the specified volume fraction range (0.01-0.95) proves challenging, one should aim to generate as much training data as possible (with two training examples having a volume fraction difference of 0.01) while being confined to a narrower range of volume fractions.

\begin{table}[h!]
	\caption{Optimized cantilever beam with different  number of neurons $n$ in the adaptive layer (Fig.~\ref{fig:CNNarchitech}). $V_\text{err}$ and $Obj_\text{err}$ indicate the volume and objective error between the results obtained by the proposed CNN and the target output (generated by MATLAB codes).} \label{Table:CNNresults}
\resizebox{\textwidth}{!}{
\begin{tabular}{|l|l|cccccc|l|}
\hline
\textbf{$V_f$}        & \textbf{Input}                                        & \multicolumn{6}{c|}{\textbf{CNN results}}                                                                                                                                                                                                                                                                                    & \textbf{Target}                  \\ \hline
\multirow{5}{*}{0.03} & \multirow{5}{*}{\vspace{-1cm}\includegraphics[scale=0.075]{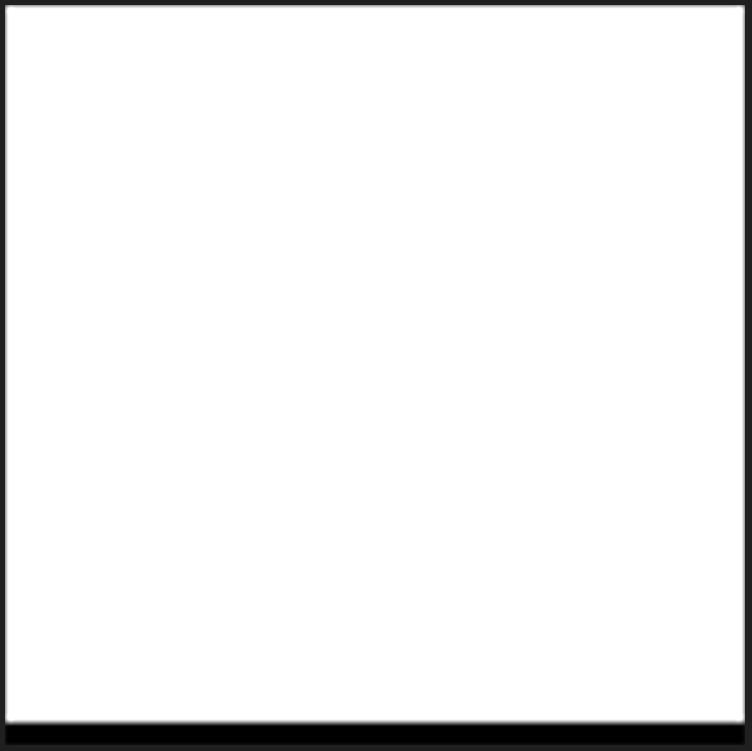}}                      & \multicolumn{6}{c|}{\textbf{\# neurons in adaptive layer ($n$)}}                                                                                                                                                                                                                                                         & \multirow{5}{*}{\vspace{-1cm}\includegraphics[scale=0.5]{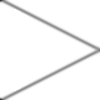}} \\ \cline{3-8}
                      &                                                       & \multicolumn{1}{l|}{\textbf{1000}}                    & \multicolumn{1}{l|}{\textbf{2000}}                    & \multicolumn{1}{l|}{\textbf{4000}}                    & \multicolumn{1}{l|}{\textbf{8000}}                    & \multicolumn{1}{l|}{\textbf{12000}}                   & \multicolumn{1}{l|}{\textbf{16000}}  &                                  \\ \cline{3-8}
                      &                                                       & \multicolumn{1}{l|}{\includegraphics[scale=0.25]{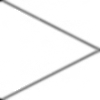}}                  & \multicolumn{1}{l|}{\includegraphics[scale=0.25]{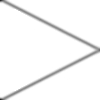}}                  & \multicolumn{1}{l|}{\includegraphics[scale=0.25]{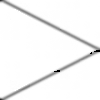}}                  & \multicolumn{1}{l|}{\includegraphics[scale=0.25]{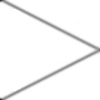}}                  & \multicolumn{1}{l|}{\includegraphics[scale=0.25]{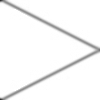}}                  & \multicolumn{1}{l|}{\includegraphics[scale=0.25]{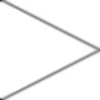}} &                                  \\ \cline{3-8}
                      &                                                       & \multicolumn{6}{c|}{\{$V_\text{err}$ , $Obj_\text{err}$\} (\%)}                                                                                                                                                                                                                                                                   &                                  \\ \cline{3-8}
                      &                                                       & \multicolumn{1}{l|}{\{3.3, 2.76\}}                    & \multicolumn{1}{l|}{\{17, 7.9\}}                      & \multicolumn{1}{l|}{\{10.3, 15.02\}}                  & \multicolumn{1}{l|}{\{1,  2.48\}}                     & \multicolumn{1}{l|}{\{1, 2.76\}}                      & \multicolumn{1}{l|}{\{0.33, 0.89\}}  &                                  \\ \hline
\multirow{6}{*}{0.25} & \multirow{6}{*}{\includegraphics[scale=0.075]{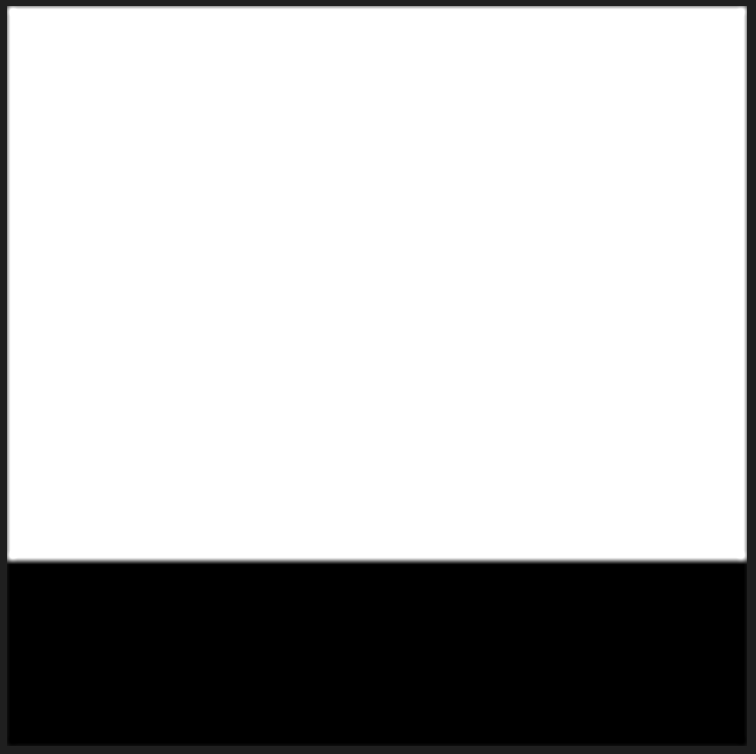}}                      & \multicolumn{1}{c|}{\multirow{4}{*}{\includegraphics[scale=0.25]{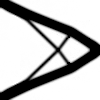}}} & \multicolumn{1}{c|}{\multirow{4}{*}{\includegraphics[scale=0.25]{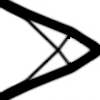}}} & \multicolumn{1}{c|}{\multirow{4}{*}{\includegraphics[scale=0.25]{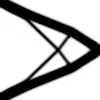}}} & \multicolumn{1}{c|}{\multirow{4}{*}{\includegraphics[scale=0.25]{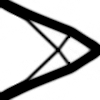}}} & \multicolumn{1}{c|}{\multirow{4}{*}{\includegraphics[scale=0.2]{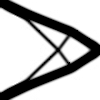}}} & \multirow{4}{*}{\includegraphics[scale=0.2]{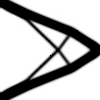}}     & \multirow{6}{*}{\includegraphics[scale=0.5]{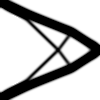}} \\
                      &                                                       & \multicolumn{1}{c|}{}                                 & \multicolumn{1}{c|}{}                                 & \multicolumn{1}{c|}{}                                 & \multicolumn{1}{c|}{}                                 & \multicolumn{1}{c|}{}                                 &                                      &                                  \\
                      &                                                       & \multicolumn{1}{c|}{}                                 & \multicolumn{1}{c|}{}                                 & \multicolumn{1}{c|}{}                                 & \multicolumn{1}{c|}{}                                 & \multicolumn{1}{c|}{}                                 &                                      &                                  \\
                      &                                                       & \multicolumn{1}{c|}{}                                 & \multicolumn{1}{c|}{}                                 & \multicolumn{1}{c|}{}                                 & \multicolumn{1}{c|}{}                                 & \multicolumn{1}{c|}{}                                 &                                      &                                  \\ \cline{3-8}
                      &                                                       & \multicolumn{6}{c|}{\{$V_\text{err}$ , $Obj_\text{err}$\} (\%)}                                                                                                                                                                                                                                                          &                                  \\ \cline{3-8}
                      &                                                       & \multicolumn{1}{l|}{\{0.6, 0.71\}}                    & \multicolumn{1}{l|}{\{0.12, 0.45\}}                   & \multicolumn{1}{l|}{\{0.12, 0.31\}}                   & \multicolumn{1}{l|}{\{0.44, 0.41\}}                   & \multicolumn{1}{l|}{\{0.024, 0.34\}}                  & \multicolumn{1}{l|}{\{0.6, 0.56\}}   &                                  \\ \hline
\multirow{6}{*}{0.50} & \multicolumn{1}{c|}{\multirow{6}{*}{\includegraphics[scale=0.075]{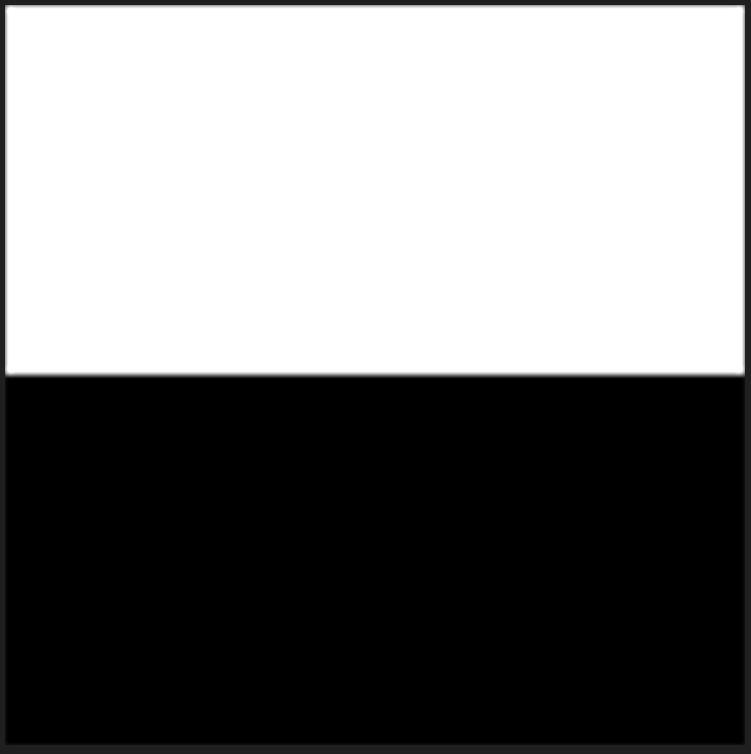}}} & \multicolumn{1}{c|}{\multirow{4}{*}{\includegraphics[scale=0.25]{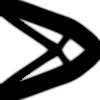}}} & \multicolumn{1}{c|}{\multirow{4}{*}{\includegraphics[scale=0.25]{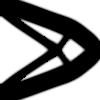}}} & \multicolumn{1}{c|}{\multirow{4}{*}{\includegraphics[scale=0.2]{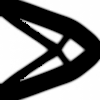}}} & \multicolumn{1}{c|}{\multirow{4}{*}{\includegraphics[scale=0.25]{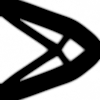}}} & \multicolumn{1}{c|}{\multirow{4}{*}{\includegraphics[scale=0.25]{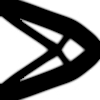}}} & \multirow{4}{*}{\includegraphics[scale=0.2]{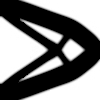}}     & \multirow{6}{*}{\includegraphics[scale=0.35]{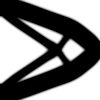}} \\
                      & \multicolumn{1}{c|}{}                                 & \multicolumn{1}{c|}{}                                 & \multicolumn{1}{c|}{}                                 & \multicolumn{1}{c|}{}                                 & \multicolumn{1}{c|}{}                                 & \multicolumn{1}{c|}{}                                 &                                      &                                  \\
                      & \multicolumn{1}{c|}{}                                 & \multicolumn{1}{c|}{}                                 & \multicolumn{1}{c|}{}                                 & \multicolumn{1}{c|}{}                                 & \multicolumn{1}{c|}{}                                 & \multicolumn{1}{c|}{}                                 &                                      &                                  \\
                      & \multicolumn{1}{c|}{}                                 & \multicolumn{1}{c|}{}                                 & \multicolumn{1}{c|}{}                                 & \multicolumn{1}{c|}{}                                 & \multicolumn{1}{c|}{}                                 & \multicolumn{1}{c|}{}                                 &                                      &                                  \\ \cline{3-8}
                      & \multicolumn{1}{c|}{}                                 & \multicolumn{6}{c|}{\{$V_\text{err}$ , $Obj_\text{err}$\} (\%)}                                                                                                                                                                                                                                               &                                  \\ \cline{3-8}
                      & \multicolumn{1}{c|}{}                                 & \multicolumn{1}{l|}{\{0.22, 0.056 \}}                 & \multicolumn{1}{l|}{\{0.04, 0.28\}}                   & \multicolumn{1}{l|}{\{0.72, 0.28\}}                   & \multicolumn{1}{l|}{\{0.28, 0.037\}}                  & \multicolumn{1}{l|}{\{0.08, 0.09\}}                   & \multicolumn{1}{l|}{\{0.16, 0.14\}}  &                                  \\ \hline
\multirow{6}{*}{0.75} & \multicolumn{1}{c|}{\multirow{6}{*}{\includegraphics[scale=0.075]{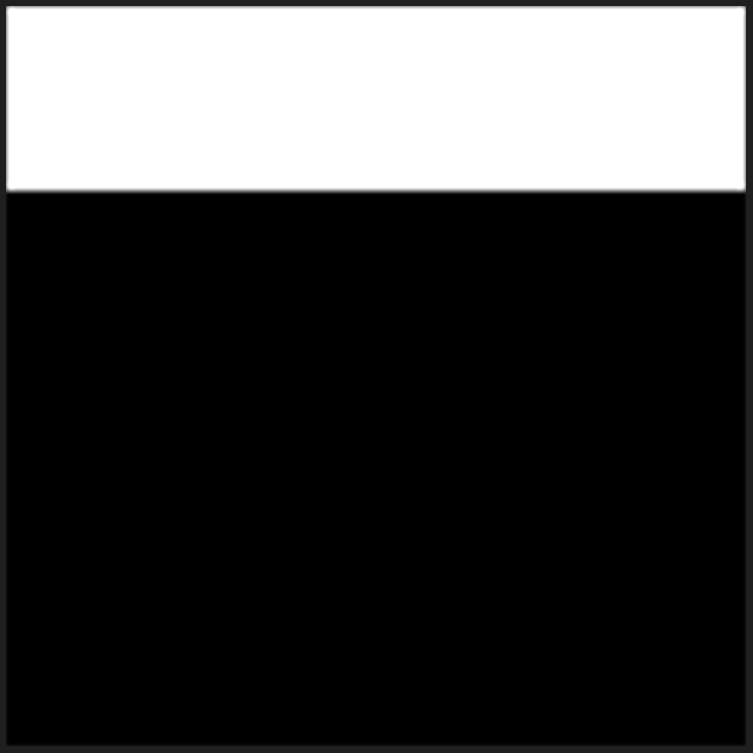}}} & \multicolumn{1}{c|}{\multirow{4}{*}{\includegraphics[scale=0.2]{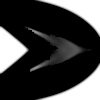}}} & \multicolumn{1}{c|}{\multirow{4}{*}{\includegraphics[scale=0.25]{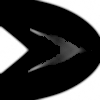}}} & \multicolumn{1}{c|}{\multirow{4}{*}{\includegraphics[scale=0.25]{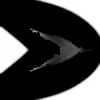}}} & \multicolumn{1}{c|}{\multirow{4}{*}{\includegraphics[scale=0.25]{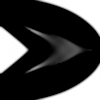}}} & \multicolumn{1}{c|}{\multirow{4}{*}{\includegraphics[scale=0.25]{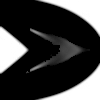}}} & \multirow{4}{*}{\includegraphics[scale=0.25]{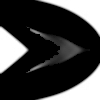}}     & \multirow{6}{*}{\includegraphics[scale=0.5]{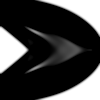}} \\
                      & \multicolumn{1}{c|}{}                                 & \multicolumn{1}{c|}{}                                 & \multicolumn{1}{c|}{}                                 & \multicolumn{1}{c|}{}                                 & \multicolumn{1}{c|}{}                                 & \multicolumn{1}{c|}{}                                 &                                      &                                  \\
                      & \multicolumn{1}{c|}{}                                 & \multicolumn{1}{c|}{}                                 & \multicolumn{1}{c|}{}                                 & \multicolumn{1}{c|}{}                                 & \multicolumn{1}{c|}{}                                 & \multicolumn{1}{c|}{}                                 &                                      &                                  \\
                      & \multicolumn{1}{c|}{}                                 & \multicolumn{1}{c|}{}                                 & \multicolumn{1}{c|}{}                                 & \multicolumn{1}{c|}{}                                 & \multicolumn{1}{c|}{}                                 & \multicolumn{1}{c|}{}                                 &                                      &                                  \\ \cline{3-8}
                      & \multicolumn{1}{c|}{}                                 & \multicolumn{6}{c|}{\{$V_\text{err}$ , $Obj_\text{err}$\} (\%)}                                                                                                                                                                                                                                                                   &                                  \\ \cline{3-8}
                      & \multicolumn{1}{c|}{}                                 & \multicolumn{1}{l|}{\{0.08, 0.13\}}                   & \multicolumn{1}{l|}{\{0.41, 0.47\}}                   & \multicolumn{1}{l|}{\{0.6, 0.68\}}                    & \multicolumn{1}{l|}{\{0.0, 0.01\}}                    & \multicolumn{1}{l|}{\{0.34, 0.34\}}                   & \multicolumn{1}{l|}{\{0.36, 0.41\}}  &                                  \\ \hline
\end{tabular}}
\end{table}

%%%%%%%%%%%%%%%%%%%%%%%%%%%%%%%%%%%%%%%%%%%%%%%%%%%%

\section{Results and discussions}\label{Sec:Sec3_res_dis}
This section demonstrates  efficacy of the proposed architecture on problems involving three different physics (Fig.~\ref{fig:Training set}) used for training the architecture. 

\subsection{Cantilever beam problem}
The cantilever beam design domain (Fig.~\ref{fig:cantileverDD}) is solved herein. First, the \textit{Base} architecture is trained for 2000 epochs with the data generated per Sec.~\ref{Sec:GentrainingData}. However, this network's optimized designs (output) did not meet the desired quality. These designs feature intricate patterns, and the \textit{Base} architecture lacked sufficient trainable parameters to capture the underlying patterns embedded within the training data effectively. To enhance the model learning capabilities, we introduce an adaptive layer containing $n$ neurons within the dense layers (Fig.~\ref{fig:CNNarchitech}). Users can change $n$ as per the complexity of the complexity of the optimized geometrical features.

Table~\ref{Table:CNNresults} depicts the optimized results with different $n$. Errors in volume and performance are denoted via $\{V_\text{err},\,obj_\text{err}\}$ in the table.  $\{V_\text{err},\,obj_\text{err}\}$ with  $n=8000$ are low, which shows the optimized designs obtained by the proposed CNN are in perfect agreement with those generated via the MATLAB code. However, using $n=8000$ in the adaptive layer may not always give low errors for the different optimization problems. To demonstrate the diversity and capability of the proposed network, next, we present design-dependent loadbearing arch structure~\cite{kumar2020topology,kumar2023TOPress} and material bulk modulus optimization problems~\cite{xia2015design}.

\begin{figure}[h!]
\centering
	\begin{subfigure}{0.45\textwidth}
		\centering
		\includegraphics[scale=0.15]{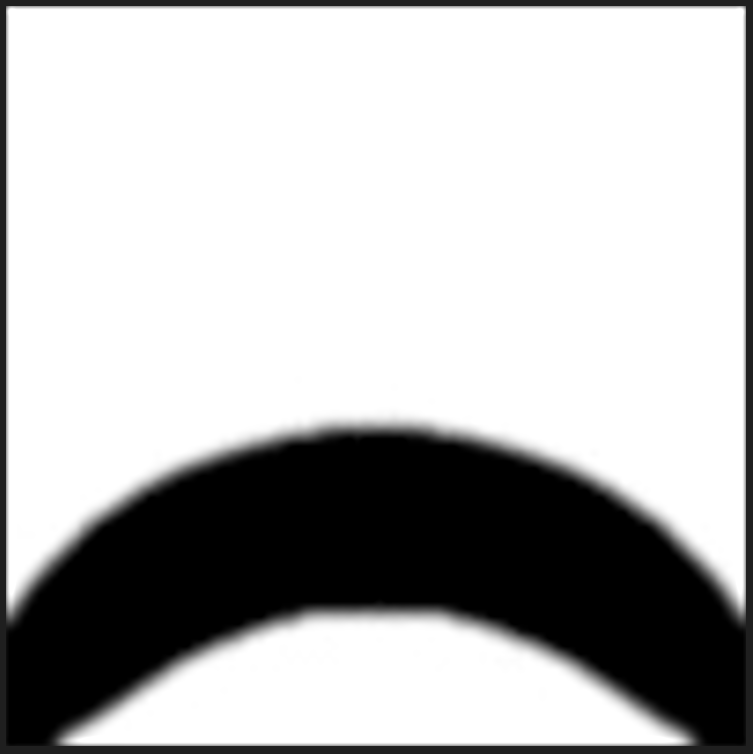}
		\caption{$V_\text{err} = 0.96\%$}		\label{fig:press_vf_25_final}
	\end{subfigure}
\quad
 	\begin{subfigure}{0.45\textwidth}
		\centering
		\includegraphics[scale=0.15]{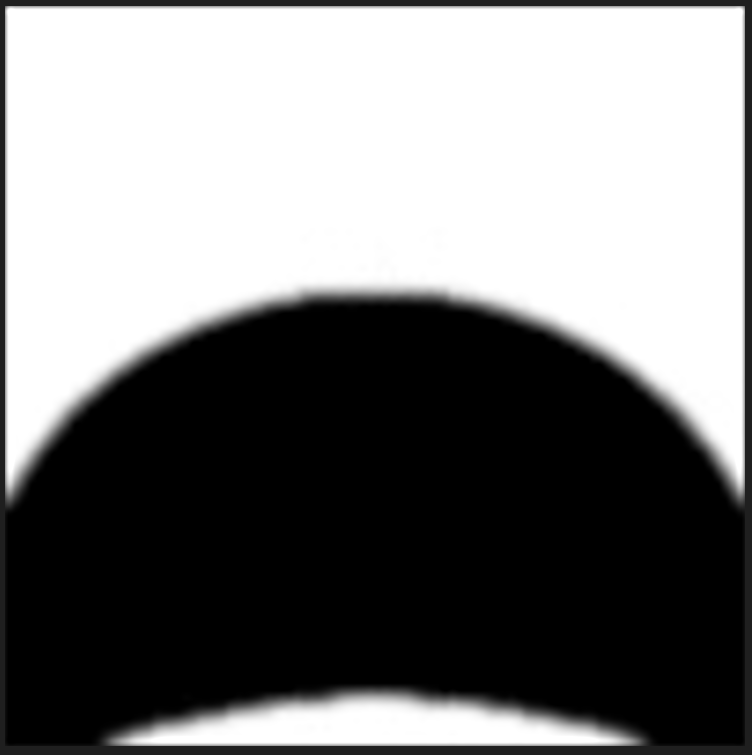}
		\caption{$V_\text{err} = 0.1\%$}\label{fig:press_vf_50_final}
	\end{subfigure}
	\caption{Optimized loadbearing arch structures at different volume fractions. (\subref{fig:press_vf_25_final}) $V_f = 0.25$ (\subref{fig:press_vf_50_final}) $V_f = 0.50$. $V_\text{err}$ indicates the volume fraction errors with respect to the target results.} \label{fig:TOPress_results_CNN}
\end{figure}

\begin{figure}[h!]
\centering
	\begin{subfigure}{0.45\textwidth}
		\centering
		\includegraphics[scale=0.15]{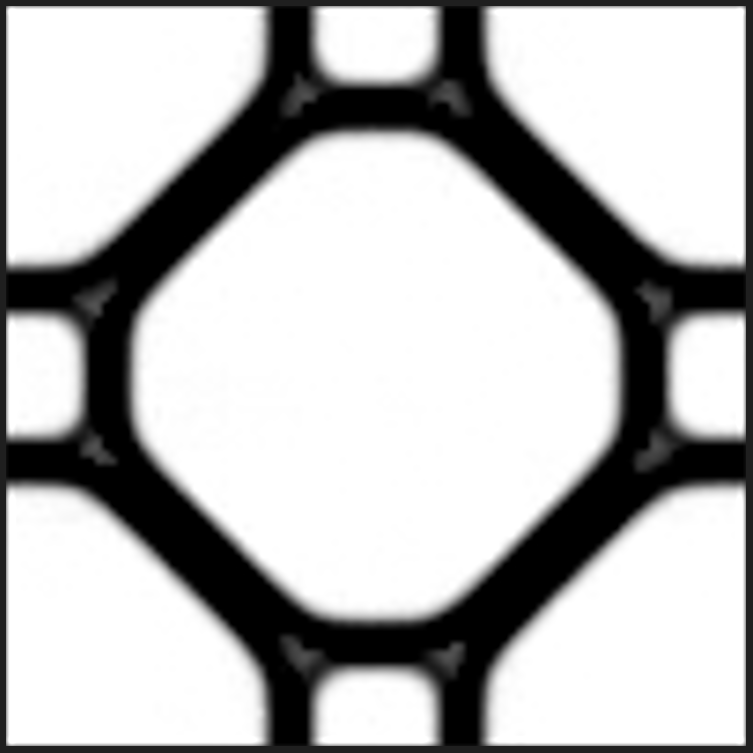}
		\caption{$V_\text{err} = 0.84\%$}		\label{fig:topX-vf-0.25}
	\end{subfigure}
\quad
 	\begin{subfigure}{0.45\textwidth}
		\centering
		\includegraphics[scale=0.15]{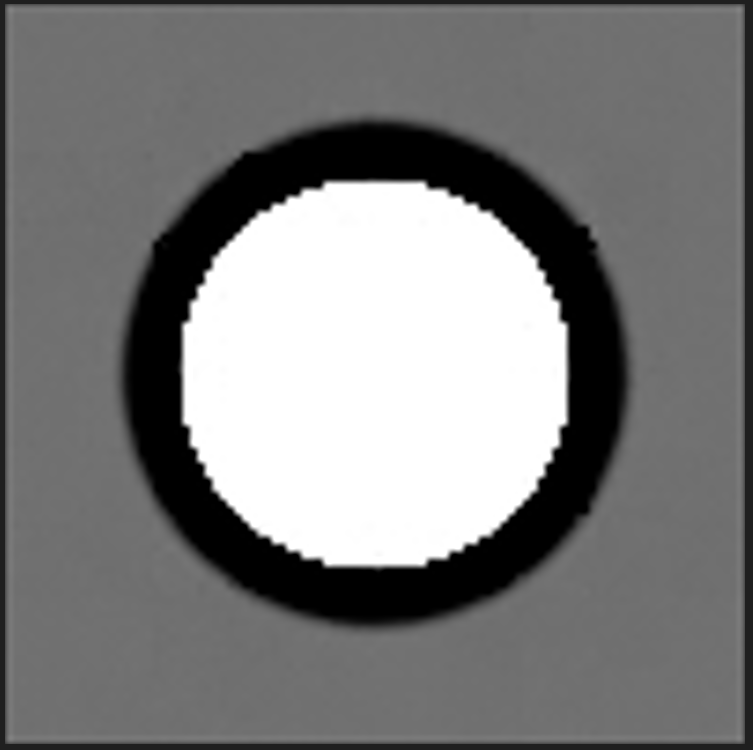}
		\caption{$V_\text{err} = 0.1\%$}\label{fig:topX-vf-0.50}
	\end{subfigure}
	\caption{Results for material bulk modulus optimization for different volume fractions. (\subref{fig:topX-vf-0.25})$V_f = 0.25$ (\subref{fig:topX-vf-0.50}) $V_f = 0.50$. } \label{fig:topX_results_CNN}
\end{figure}

\subsection{Pressure loadbearing arch structure}
The design domain is depicted in Fig.~\ref{fig:archDD} for pressure loadbearing arch structure. Though the physics involved is complex~\cite{kumar2020topology,kumar2023TOPress}, the obtained optimized arch geometrical feature is simple. Therefore, $n=0$ is used, i.e., the adaptive layer is not used in this case while training the proposed network. 

Figure~\ref{fig:TOPress_results_CNN} shows the optimized loadbearing arch designs obtained for different volume fractions ($V_f$) using the proposed neural network. The output results resemble those obtained via the \texttt{TOPress} code~\cite{kumar2023TOPress} with less than $1\%$ $V_\text{err}$, i.e., negligible $V_\text{err}$. Developing such a network for 3D design-dependent problems~\cite{kumar2021topology} and pneumatically activated soft robots~\cite{kumar2023sorotop} can be one of the research directions.

\subsection{Material bulk modulus optimization problem}
Next, the material bulk modulus optimization problem is solved (Fig.~\ref{fig:microDD}). This is considered one of the involved problems in TO~\cite{xia2015design}. 

Noting the complexity in the optimized designs, the adaptive layer with $n=4000$ is employed while training the network. Outputs of the network with different volume fractions are depicted in Fig.~\ref{fig:topX_results_CNN}. The volume fraction errors are negligible.

\section{Concluding remarks}\label{Sec:Sec4_Conclusion}
This paper proposes an adaptive deep  Convolutional Neural Network approach to tackle the TO problem. The architect employs an encoder-decoder network with dense layers wherein an adaptive layer with $n$ neurons is introduced to help capture the complex geometrical features for the optimized designs. Users can vary $n$ as per the desired level of accuracy. Three publicly available MATLAB codes are used to generate the data for training purposes. The efficacy and success of the developed CNN architecture are tested by generating optimized designs for compliance minimization problems with constant and design-dependent loads and material bulk modulus optimization. Once the network is trained with a certain number of epochs, it gives the sought-optimized designs instantaneously with negligible error in volume w.r.t. to the target designs. 

The proposed model has a fixed domain of $100\times100$, and output (optimized designs) can be generated explicitly for the boundary and force conditions for which the training data has been provided. Tapping the power of the deep neural network, generalizing the proposed network for design domain size, and obtaining the output results for the boundary and force conditions that are not used for the training data open up exciting avenues for further research.

\end{document}